FERMILAB-CONF-18-172-AD

# OVERVIEW OF FABRICATION TECHNIQUES AND LESSONS LEARNED WITH ACCELERATOR VACUUM WINDOWS*

C.R. Ader[†], M.W. McGee, L.E. Nobrega and E.A. Voirin,
Fermi National Accelerator Laboratory, Batavia, IL 60510, USA*Abstract*

Vacuum thin windows have been used in Fermilab's accelerators for decades and typically have been overlooked in terms of their criticality and fragility. Vacuum windows allow beam to pass through while creating a boundary between vacuum and air or high vacuum and low vacuum areas. The design of vacuum windows, including Titanium and Beryllium windows, will be discussed as well as fabrication, testing, and operational concerns. Failure of windows will be reviewed as well as safety approaches to mitigating failures and extending the lifetimes of vacuum windows. Various methods of calculating the strengths of vacuum windows will be explored, including FEA.## INTRODUCTION

A vacuum window is any relatively thin separation between a volume under vacuum and a volume at atmospheric pressure or vacuum through which primary or secondary beam passes. However, a thin window can also be a thin separation between atmosphere and a pressurized gas. This definition does not include optical windows. Low mass thin vacuum windows are required to minimize beam loss in the beam lines of particle accelerators and are essentially used to allow to place instrumentation into the beam line or are used at the end of beam lines before a target or beam absorber. They exhibit a vacuum integrity which allows them to be used in vacuum systems with a pressure of $10^{-10}$ Torr.

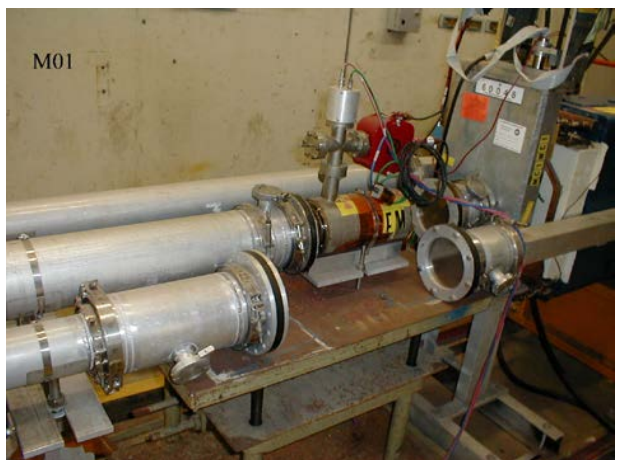

Figure 1: Example of vacuum windows in SY120 (picture taken in 03/2005).

___________________________________________
* Operated by Fermi Research Alliance, LLC, under Contract No. DE-AC02-07CH11359 with the U.S. Department of Energy.
† cader@fnal.govVacuum windows are typically fabricated of Titanium, but other materials have been used, including Beryllium. One of the main factors in deciding what material to use is the Z-factor, due to its low atomic number and transparency to energetic particles.

The Tevatron (KTeV) Kevlar-Mylar vacuum window was the largest vacuum window fabricated at Fermilab and had 1.8 m (diameter aperture) components. The KTeV decay pipe window had a factor of safety = 1.93 and building a window with a factor of safety greater than 2.0 was not possible or would have prevented the experiment from meeting its physics goals. So, a shutter was built and interlocked to close when the key tree was opened, effectively removing the hazard of a window failure from harming people or equipment.

## WINDOW DESIGN

Vacuum window assemblies must provide reliable mechanical performance to handle both a static differential pressure between the atmosphere and the vacuum, and occasional pressure cycling when the vacuum system is vented for maintenance. Furthermore, the window housing needs to be thin enough for beam requirements.

Roark's and Young's analytical solutions are typically obtained first along with using Western's Technical Memo [1 and 2]. Then through analytical calculations and Finite Element Analysis (FEA), the calculations must show that the window will adequately withstand the vacuum load, thermal stresses due to beam passing through, and the unlikely event of internal pressure.

One FEA analysis is of the NuMI Pre-Target Beryllium window. A transient thermal finite analysis was completed of a .015in Beryllium PF60 window at one MW beam. The initial conditions showed a temperature of 38 ºC (Figure 2) and then it raised up to 98 ºC [3].

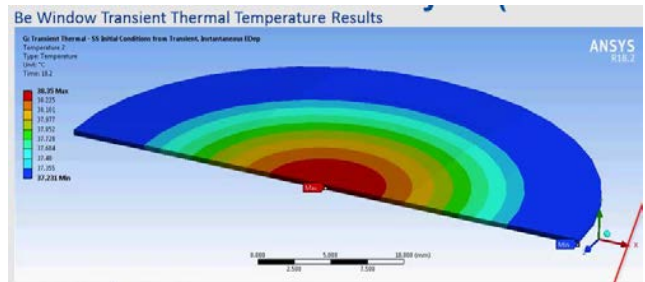

Figure 2: NuMI Pre-Target Beryllium Temperature Results.

Other analyses that should be considered are shock events, energy deposition (steady-state and transient thermal), and static structural FEA. Other environmental considerations include corrosive components.

Beryllium windows are used for their high transmission of X-rays. The material exhibits higher transmission for thinner windows. Consequently, thin vacuum windows are often used for X-ray detection systems and as a filter for lower energy particles [4].

A combined Beryllium/Titanium window assembly is preferred in some design situations which substantially reduces the possibility of spreading Beryllium contamination into the beam line from a ruptured Beryllium window [5].

## WINDOW FABRICATION

Titanium vacuum windows have been fabricated by an electron-beam process which involves first sandwiching the foil between two Titanium weldments (or rings) as shown in Figure 3. This sub-assembly is leak checked. Next, this weldment is hand welded into a custom Titanium Conflat-flange. Then the entire assembly is leak checked.

This design is symmetrical and the edges are smooth and have no sharp edges. This allows the flange to accept pressure from either side and is a much more robust design.

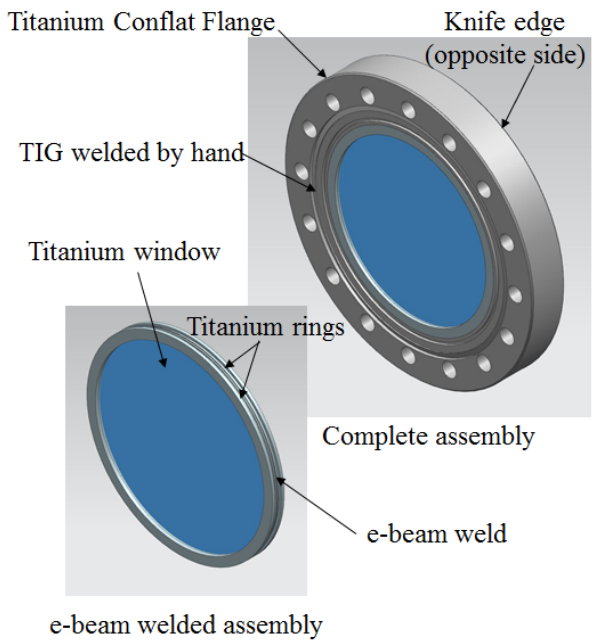

Figure 3: Titanium vacuum window assembly.

The Beryllium window assemblies are made through a two-step vacuum furnace braze. Strength is improved if the joint is between 0.0127 to 0.0381 mm (0.0254 to 0.0762 mm on the diameter). The temperature cycle for beryllium brazing is a soak at 650 ºC for 10 to 15 minutes, with a final temperature of about 800 ºC maximum for seven minutes. On a very heavy assembly, a 450 ºC soak should be added [6].

Due to the brittleness of the thin Beryllium disk and the braze joint, extreme care must be taken to avoid sudden pressure changes across the window aperture, either during pumping down or venting to atmospheric pressure. This is especially the case for the first pumping down of the window assembly after the second braze cycle. To avoid a failure, the brazed window assembly should be evacuated in a very controlled fashion to slowly form a stable curved state.

The Fermilab NuMI Beryllium window had an initial deformation of the window of approximately 0.508 mm. This deformation comes from the brazing operation of the manufacturing process. [5]. Typically, the dishing or performing of a window reduces the stresses for a given load. Often, a flat window will take a set with some deflection after initial pump down.

The final design consideration is a result of maintaining the device over it's time in service. This will require technicians to let the device up to atmospheric pressure from vacuum. Infrequent mechanical cycling due to vacuum system purging with an inert gas at a 1.103 Bar (16 psia positive pressure or 1 psig given a poppet style safety relief on the vent line) for maintenance reasons represents only a finite amount of cycles over each window's lifetime. This will likely be done less than five times through the life of the device. Therefore, the fatigue strength of the device will not be considered. In the case of Grade 5 Titanium, the fatigue data available is on the order of $10^7$ cycles. Therefore, fatigue is not an issue regarding this Titanium window.

## WINDOW FAILURE MODES

A window failure of any design may take several forms from conception to commissioning, such as the example shown in Figure 4. The contributing factors may include: lack of consistent fabrication, human error, cultural safety issues within an organization, inadequate oversight and line-management issues. Examples involve the acute and chronic failures of Fermilab's Muon Test Area (MTA) and NuMI windows, respectively.

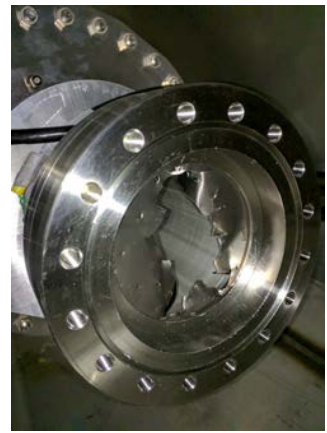

Figure 4: Titanium vacuum window failure.

In the design process, consider a Titanium catcher window for a Beryllium window. The consequences of Beryllium contamination can be severe and can include inhalation of Beryllium dust and/or particles which can cause Chronic Beryllium Disease (CBD) or Beryllium sensitization in exposed individuals. CBD is a chronic and sometimes fatal lung condition. Beryllium sensitization is a condition in which a person's immune system may become highly responsive or allergic to the presence of any Beryllium within the body.

The NuMI beam line Beryllium/Titanium window assembly shown in Figure 5 was located at the most downstream end of the 3.35 m beam line. At this location, after passing through a 2.1 m thick shield wall, the beam line vacuum terminates with the vacuum assembly. The Beryllium window maintains a vacuum of $1 \times 10^{-7}$ Torr.

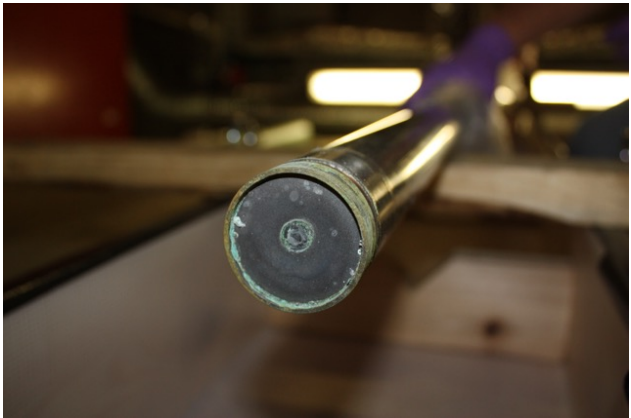

Figure 5: NuMI beam line Beryllium vacuum window

This window failed after seven years and $7 \times 10^7$ pulses and developed a small $10^{-8}$ permeation vacuum leak in the outer braze area most likely due to corrosion from the humid and nitric acid environment. There is a greenish copper oxide on most of the outer braze joint. But the pin hole leak can be seen on the outer diameter with a whitish corrosion surrounding it. The Computerized Tomography (CT) scan shows clearly that the window braze-joint failure initiated the Beryllium contamination.

## OPERATIONAL CONCERNS

A critical component of all vacuum systems is the relief device. Infrequent mechanical cycling due to vacuum system purging with an inert gas at a 1.103 Bar (16 psia positive pressure or 1 psig given a poppet style safety relief on the vent line) for maintenance reasons represents only a finite amount of cycles over each window's lifetime. In the case of Grade 5 Titanium, the fatigue data available is on the order of $10^7$ cycles. Therefore, fatigue is not an issue regarding these Titanium windows.

The blue line shown in Figure 6 represents the 0.5 MeV-cm$^2$/g stopping power, which is associated the absorbed energy of the beam. Using the density of the Titanium window (4.43 g/cm$^3$), the stopping power can be calculated in terms of energy deposition per unit thickness [7].

$$\frac{dE_{stopping.power}}{dx_{thickness}} = 0.5 \frac{MeV-cm^2}{g} \times 4.43 \frac{g}{cm^3} = 2.2 \frac{MeV}{cm}$$

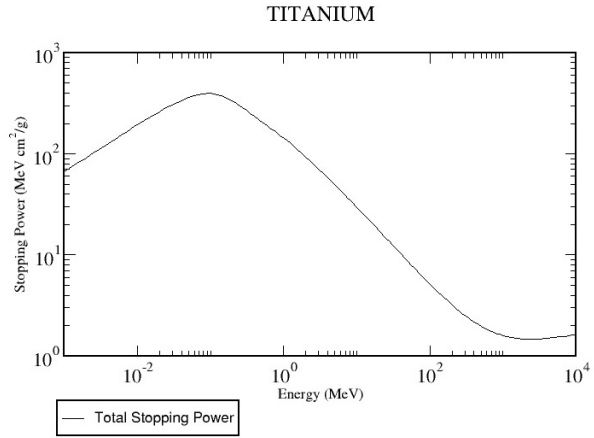

Figure 6: Titanium Stopping Power versus Beam Energy.

This value is then multiplied by the thickness of the Titanium foil, which is $5.08 \times 10^{-3}$ cm. Therefore, the 'worst case' beam energy loss through the thickness of the material is 2.2MeV/cm*$5.08 \times 10^{-3}$ cm=11.11KeV. The beam current for a single spill on a per second basis is 1.67 μA. Therefore, the total power deposited onto the surface of the foil is roughly 18.5 mW.

The final design consideration is a result of maintaining the device over its time in service. This will require technicians to let the device up to atmospheric pressure from vacuum slowly. This will likely be done no more than five times through the life of the device. Therefore, the fatigue strength of the device needs to be considered.

## FUTURE WORK

We plan on investigating manufacturing techniques to add pre-curvature to windows which will reduce the stress. Typically, a small dish in the window is produced during the brazing process from the cavity and from the leak checking process.

One company, Moxtek, that produces Dura Beryllium windows with a DuraCoat material that prevents corrosion by encapsulating the braze joint with a non-galvanic material or investigate coating the window with Parylene-C coatings for NuMI operating environment (about 8 ppm moisture and 0.5 ppm nitric acid at 40ºC) should be investigated [8].

Other areas of research should include investigating brittle versus ductile failures and comparing vacuum window typical material such as Beryllium to Titanium. The brittle failure mode can create shards which can be projected into the air. A higher Young's Modulus equates to higher stiffness. A stiffer component bends less under stress, thereby reducing mechanical deformation (breakup) and shifting resonate frequencies outside the audible range. Beryllium is 2.67 times stiffer than Titanium [4].